\documentclass[a4paper,12pt]{article}

\usepackage{amssymb,dsfont,amsmath,graphicx,esint,latexsym}

\begin{document}

\begin{center}
{\Large \bf Dark solitons of the Qiao's hierarchy }

\vskip.8cm

{\Large \it Rossen  I. Ivanov  \footnote{E-mail:
Rossen.Ivanov@dit.ie} and Tony Lyons \footnote{E-mail:
Tony.Lyons@mydit.ie}  } \vskip.8cm School of Mathematical
Sciences, Dublin Institute of Technology,

Kevin Street, Dublin 8, Ireland

\vskip0.32cm

\end{center}

\vskip1.32cm

\begin{abstract}
\noindent We obtain a class of soliton solutions of the integrable
hierarchy which has been put forward in a series of works by Z.
Qiao. The soliton solutions are in the class of real functions
approaching constant value fast enough at infinity, the so-called 'dark solitons'.  \vskip.4cm


\vskip.4cm

\end{abstract}
\section{Introduction}

The interest inspired by the Camassa-Holm (CH) equation and its
singular peakon solutions \cite{CH93}  prompted search for other
integrable equations with similar properties. An integrable peakon
equation with cubic nonlinearities has been discovered first by
Qiao \cite{Q06} and studied further e.g. in \cite{Q07,Q09}.
Another equation with cubic nonlinearities has been found by V.
Novikov \cite{N09}. The Lax pair for the Novikov's equation is
given in \cite{HJPW08}, (see also a remark on the peakons of
Qiao's equation in \cite{HJPW08}). Actually the Qiao's equation
\begin{equation}m_t+(m(u^2-u_x^2))_x=0, \qquad m=u-u_{xx} \label{Q} \end{equation}
together with the CH equation
\begin{equation}m_t+2u_xm+um_x=0, \qquad m=u-u_{xx}
\label{CH} \end{equation} belong to the bi-Hamiltonian hierarchy
of equations described by Fokas and Fuchssteiner \cite{FF80}. The
Qiao's equation has a distinctive $W/M$-shape travelling wave
solutions \cite{Q06,Q07}. The peakons of Novikov's equation have
been studied in \cite{HLS09}. $2+1$ dimensional generalizations of
Qiao's hierarchy are studied in \cite{E11}. Single peakon,
mutil-peakon dynamics, weak kink, kink-peakon, and stability
analysis of the Qiao's equation were studied in \cite{QXL12} and
\cite{GLOQ12}. For the CH and related equations one can consult
the monographs \cite{HSS2009,AC11,PS11} and the references
therein.

Equation (\ref{Q}) can also be written as
\begin{equation}\label{qiao2} m_t + (u^2-u_x^2)m_x +2u_xm^2=0.
\end{equation}

Qiao presented a $2\times 2$  Lax pair for this equation given by
the linear system ${\bf \Psi}_x={\bf U} {\bf \Psi}$, ${\bf \Psi}_t
= {\bf V} {\bf \Psi}$ with \begin{eqnarray}\label{qlax}
\begin{array}{ccl}
{\bf U} & = &  \left( \begin{array}{cc}
-\frac{1}{2}& \frac{1}{2}m \lambda \\
-\frac{1}{2}m \lambda & \frac{1}{2}
\end{array} \right),
\\
{\bf V} &= & \left( \begin{array}{ccc}
\lambda^{-2}+  \frac{1}{2}(u^2-u_x^2)&-\lambda^{-1}(u-u_x)-\frac{1}{2}m\lambda (u^2-u_x^2)\\
 \lambda^{-1}(u+u_x)+\frac{1}{2}m\lambda (u^2-u_x^2) &
-\lambda^{-2}-  \frac{1}{2}(u^2-u_x^2)
\end{array} \right)   .
\end{array}
\end{eqnarray}

\noindent There is another equation from the same hierarchy,
\begin{equation}\label{Q2} m_t + \left(\frac{1}{m^2}\right)_x -\left(\frac{1}{m^2}\right)_{xxx}=0,
\end{equation} for which the ${\bf V}$-operator is ${\bf \Psi}_t
= {\bf V_2} {\bf \Psi}$ where

\begin{eqnarray}\label{V2}
{\bf V_2} &= & \frac{\lambda}{2}\left( \begin{array}{ccc}
-\frac{\lambda}{m}&\lambda^{2}+\frac{m(m_x-m_{xx})+3m_x^2}{m^4}\\
-\lambda^{2}+\frac{m(m_x+m_{xx})-3m_x^2}{m^4} & \frac{\lambda}{m}
\end{array} \right)   .
\end{eqnarray}

The (white) soliton solutions of (\ref{Q}) and (\ref{Q2}) have
been found previously \cite{S11,ZQ}. These studies rely on the
fact that the spectral problem for (\ref{Q}) is gauge-equivalent
to the one for the mKdV equation. In this study we will present
soliton solutions approaching a constant value for $|x|\to \infty$
(dark solitons). To this end we are going to formulate the
spectral problem in the form of a Schr\"odinger operator, which is
the same spectral problem as for the KdV equation.

\section{Reformulation of the spectral problem}

Let us consider solutions such as \begin{equation} \label{m0}
m(x,t)>0, \qquad \lim_{x\to \pm \infty}m(x,t)=m_0,
\end{equation}
where $m_0$ is a positive constant. Let us assume also that
$m(x,\cdot)-m_0 \in \mathcal{S}(\mathbb{R})$ for any value of $t$.
One can reformulate the spectral problem into a scalar one as
follows. Introducing ${\bf \Psi}=(\psi, \phi)^T $ the matrix Lax
pair written in components is
\begin{eqnarray*}
    2\psi_x = -\psi + m\lambda\phi\\
    2\phi_x = -m\lambda\psi + \phi.
\end{eqnarray*}
With a change of coordinates
\begin{equation}\label{change}
\partial_y = \frac{2}{m}\partial_x, \qquad \psi=\frac{1}{\lambda}\left[\frac{\phi}{m} - \phi_y\right] \end{equation} we obtain the following scalar spectral problem for $\phi(y,\lambda)$ (sometimes we do not write the argument $t$
which is an external parameter for the considered spectral
problem)
\begin{equation}\label{Schr}
-\phi_{yy} + \left[\left(\frac{1}{m}\right)_y +
\frac{1}{m^2}\right]\phi = \lambda^2\phi.
\end{equation}
Note that this is a Schr\"odinger's operator with a potential
\begin{equation} \label{u}
U(y,t)=\left(\frac{1}{m}\right)_y + \frac{1}{m^2} \end{equation}
It is well known how to recover $U(y,t)$ from the scattering data
of (\ref{Schr}), however the solution is $m(y,t)$ and its recovery
from $U(y,t)$ necessitates solving a nonlinear (Riccati) equation.
We can express $m(y,t)$ in terms of the eigenfunctions of the
Schr\"odinger's operator. We introduce $\rho(y,\lambda) =
\frac{\phi_y}{\phi}$ from which we immediately obtain
\[\rho_y + \rho^2 = \frac{\phi_{yy}}{\phi} = U(y) - \lambda^2.\]
If we define
\[\rho_0(y) = \rho(y,0)\]
then we have
\[U(y) = \rho_{0,y} + \rho_0^2.\]
However, due to (\ref{u}) we now have $ \frac{1}{m} = \rho_0$ or
\begin{equation} m(y,t) = \frac{1}{\rho_0(y,t)} =
\left.\frac{\phi(y,t ,\lambda)}{\phi_y(y,t ,\lambda)}\right|_{\lambda = 0}
\label{my}
\end{equation}

So far we treated $y$ as a new variable instead of $x$. However we
can treat $y$ as a parameter, and then (\ref{my}) represents the
solution in parametric form, where the original variable $x$ is
given due to (\ref{change}), (\ref{my}) by:

\begin{equation} x(y,t) = 2 \ln \phi(y,t,0) +\text{const}.
\label{xy}
\end{equation}

Assuming that $\phi(y,t,0)$ is everywhere positive, we have a
solution in parametric form (\ref{my}), (\ref{xy}) given entirely
in terms of the eigenfunctions $\phi(y,t,0)$. One can write
formally the solution (neglecting the constant in the last
formula) as

\begin{equation} m(x,t) = 2 \int_{-\infty}^{\infty} \delta\left(x-2 \ln \phi(y,t,0) \right)\text{d}y. \label{mx}
\end{equation}

\section{Inverse scattering and Soliton solutions}

From (\ref{m0}) and (\ref{u}) it follows that $U(y)$ does not
decay to $0$ when $y\to \pm \infty $. To this end we introduce the
modified potential
\begin{equation} \label{umodified} \tilde{U}(y)= U(y) - \frac{1}{m_0^2},\end{equation}
for which $\lim_{|y| \to \infty} \tilde{U}(y) = 0$. So we have
\[-\phi_{yy} + \left[U(y) - \frac{1}{m_0^2}\right]\phi = \left(\lambda^2 - \frac{1}{m_0^2}\right)\phi,\]
or, introducing a new spectral parameter \begin{equation}
\label{k} k^2= \lambda^2 - \frac{1}{m_0^2}\end{equation} we have a
standard spectral problem \begin{equation}\label{Schr2} -\phi_{yy}
+ \tilde{U}(y)\phi(k,y) = k^2\phi(k,y), \qquad \tilde{U}(y)\in
\mathcal{S}(\mathbb{R}).\end{equation}
 When $\lambda = 0$ however we find
$k = \pm\frac{i}{m_0}$ for $k$. This means that if one takes an
eigenfunction $\phi(k,y)$ of (\ref{Schr2}) analytic in the upper
(lower) half complex $k$-plane, one should evaluate it at $k =
\frac{i}{m_0}$ ($k = -\frac{i}{m_0}$):

\begin{eqnarray} m(y,t)& = &
\left.\frac{\phi(y,t ,k)}{\phi_y(y,t ,k)}\right|_{k
=\pm\frac{i}{m_0}} \label{sol1}
\\
x(y,t)& = & 2 \ln \phi\left(y,t,\pm \frac{i}{m_0}\right). \label{sol2}
\end{eqnarray}

The spectral theory for the problem (\ref{Schr2}) is well
developed, e.g. \cite{ZMNP}. We are going to use these results to
construct the soliton solutions of (\ref{Q}), (\ref{Q2}). One can
introduce scattering data as usual. For the time-dependence of the
scattering data one needs the time-evolution of the eigenfunction
$\phi(k,x)$. The Lax-pair in $x$ and $t$ variables  for (\ref{Q})
has the form \begin{eqnarray}
\phi_{xx}&=&\frac{m_x}{m}\phi_{x}+\left(\frac{1}{4}-\frac{m_x}{2m}-\frac{m^2}{4}\lambda^2
\right)\phi, \label{1}
\\
\phi_t &= &
\frac{1}{\lambda^2}\left[\frac{u_x+u_{xx}}{m}\right]\phi -
\left[\frac{u+u_x}{\lambda^2m} + \frac{u^2 -
u_{x}^{2}}{2}\right]\phi_x + \gamma\phi, \label{2}
\end{eqnarray}
where $\gamma$ is an arbitrary constant. The second equation,
(\ref{2}) in asymptotic form $x\to \pm \infty$ is \begin{eqnarray}
\phi_t &\to & - \left[\frac{1}{\lambda^2} +
\frac{m_0^2}{2}\right]\phi_x + \gamma\phi, \nonumber
\end{eqnarray} or, in terms of $k$, $y$-variables when $y\to
\pm\infty$,
\begin{eqnarray} \phi_t &\to & - \frac{m_0^3}{2}\left[\frac{k^2m_0^2+3}{k^2m_0^2+1}\right]\phi_y + \gamma\phi, \label{2a}
\end{eqnarray} since
\[\displaystyle{\lim_{|y|\to\infty}} m=\displaystyle{\lim_{|y|\to\infty}} u= m_0.\]
Defining Jost solutions by \begin{equation} \label{Jost} \lim_{y\to \pm \infty}\varphi_{\pm}(y,k) e^{iky}
=1, \end{equation} such that
\begin{equation} \varphi_{-}(y,k) = a(k)\varphi_{+}(y,k)+b(k) \bar{\varphi}_{+}(y,k), \qquad k\in \mathbb{R} \label{scattData}
\end{equation}
and noting that $\varphi_- \to ae^{-iky} + be^{iky}$ when $y\to
\infty$ we find from (\ref{2a})
\begin{eqnarray}
    a_t = \frac{m_0^3}{4}\left[\frac{k^2m_0^2+3}{k^2m_0^2+1}\right](ika) + \gamma a\\
    b_t = -\frac{m_0^3}{4}\left[\frac{k^2m_0^2+3}{k^2m_0^2+1}\right](ikb) + \gamma b.
\end{eqnarray}
Requiring $a_t = 0,$ we find
\[b_t = -ik\frac{m_0^3}{2}\left(\frac{k^2m_0^2+3}{k^2m_0^2+1}\right)b(k,t)\]
and thus for the scattering coefficient $r\equiv b/a$ we have
\begin{equation}
r(k,t) = r(k,0)\exp\left[
-ik\frac{m_0^3}{2}\left(\frac{k^2m_0^2+3}{k^2m_0^2+1}\right)t\right],
\end{equation}

\noindent and for the analogue on the discrete spectrum
$k=i\kappa_n$,
\begin{equation}
R_n(t)\equiv \frac{b(i\kappa_n)}{ia'(i\kappa_n)}=R_n(0)\exp\left[
\frac{\kappa_n
m_0^3(3-\kappa_n^2m_0^2)}{2(1-\kappa_n^2m_0^2)}t\right].
\end{equation}

For the equation (\ref{Q2}) the time evolution of the spectral eigenfunctions is given by \begin{equation}
\phi_t=
\frac{v}{m}\phi +
\left(\lambda^2 -v\right)\frac{2}{m}\phi_x + \gamma\phi,\qquad v=\frac{m(m_x+m_{xx})-3m_x^2}{m^4} \label{2-2}
\end{equation} and analogous considerations give \begin{eqnarray}
r(k,t) &=& r(k,0)\exp\left[
-2ik\left(\frac{1}{m_0^2}+k^2 \right)t\right], \\
R_n(t) &=& R_n(0)\exp\left[
2\kappa_n\left(\frac{1}{m_0^2}-\kappa_n^2 \right)t\right].
\end{eqnarray}

It is convenient to introduce a dispersion law for the hierarchy, which for the considered two members is

\begin{eqnarray}
f(\kappa)=\left \{\begin{array}{lll}
\frac{\kappa m_0^3(3-\kappa^2m_0^2)}{2(1-\kappa^2m_0^2)} & \quad \mbox{for eq. (\ref{Q}) }, & \nonumber \\
\frac{2\kappa}{m_0^2}\left(1-m_0^2\kappa^2 \right) & \quad \mbox{for eq. (\ref{Q2})}.& \nonumber\\
\end{array}\right.
\end{eqnarray} Then for the whole hierarchy we can write in general \begin{eqnarray}
R_n(t) &=& R_n(0)\exp\left(f(\kappa_n)t\right). \label{R}
\end{eqnarray}
For further convenience we introduce
\[\xi_n\equiv y-\frac{f(\kappa_n)}{2\kappa_n}t-\frac{1}{2\kappa_n}\ln \frac{R_n(0)}{2\kappa_n}.\]


The eigenfunctions of the spectral problem (\ref{Schr2}) are well known, see e.g. \cite{ZMNP}. In the purely $N$-soliton case the eigenfunction, analytic in the lower complex $k$-plane is the Jost solution $\varphi_+ (y,k)$ defined in (\ref{Jost}) which has the form

\begin{equation} \label{eig}
\varphi_+(y,t,k)=e^{iky}\left(1+\sum_{n=1}^{N} \frac{\Gamma_n(y,t)}{k-i\kappa_n}\right) \end{equation} with the residues $ \Gamma_n(y,t)$ satisfying a linear system $$ \Gamma_n(y,t)=i R_n(t) e^{-2\kappa_n y} \left( 1+i\sum_{m=1}^{N} \frac{\Gamma_m(y,t)}{\kappa_n+\kappa_m}\right).
$$ The time-dependence of the scattering data is given by (\ref{R}). The $N$- soliton solution then is given in parametric form by (\ref{sol1}) and (\ref{sol2}) for the eigenfunction (\ref{eig}). The condition $0<\kappa_n < m_0^{-1}$ is sufficient to ensure smoothness of the solitons.

\section{Example: One-Soliton Solution}
The one-soliton solution corresponds to one discrete eigenvalue $k_1=i\kappa_1$, where $\kappa_1$ is real, positive and $\kappa_1< m_0^{-1}$.
The eigenfunction in this case is (\ref{eig})
\begin{equation}
\varphi_+(y,t,k) = e^{iky}\left(1 +
\frac{1}{k-i\kappa_1}\cdot\frac{i R_1(t)e^{-2\kappa_1y}}{1+\frac{R_1(t)}{2\kappa_1}e^{-2\kappa_1y}}\right).
\end{equation}
Evaluated at $k = \frac{-i}{m_0}$ we find
\[\varphi_+(y,t, \frac{-i}{m_0}) = e^{\frac{y}{m_0}}\left(1 - \frac{1}{\frac{1}{m_0} + \kappa_1}\cdot\frac{R_1(t)e^{-2\kappa_1y}}{1  +\frac{R_1(t)}{2\kappa_1}e^{-2\kappa_1y}}\right).\] From (\ref{sol1}) and (\ref{sol2}) we obtain the one-soliton solutions
\begin{eqnarray}
x(y,t)&=&\frac{2y}{m_0}+ 2\ln \left(1-\frac{\kappa_1 m_0 e^{-\kappa_1 \xi_1}}{(1+\kappa_1 m_0) \cosh \kappa_1 \xi_1}\right), \\
m(y,t) &=& \frac{m_0}{1 + \frac{\kappa_1^2m_0^2\text{sech}^2\kappa_1\xi_1}{1-m_0\kappa_1\tanh\kappa_1\xi_1}}.
\end{eqnarray}

\begin{figure}[ht]
\begin{center}
\includegraphics[scale=0.5,angle=0]{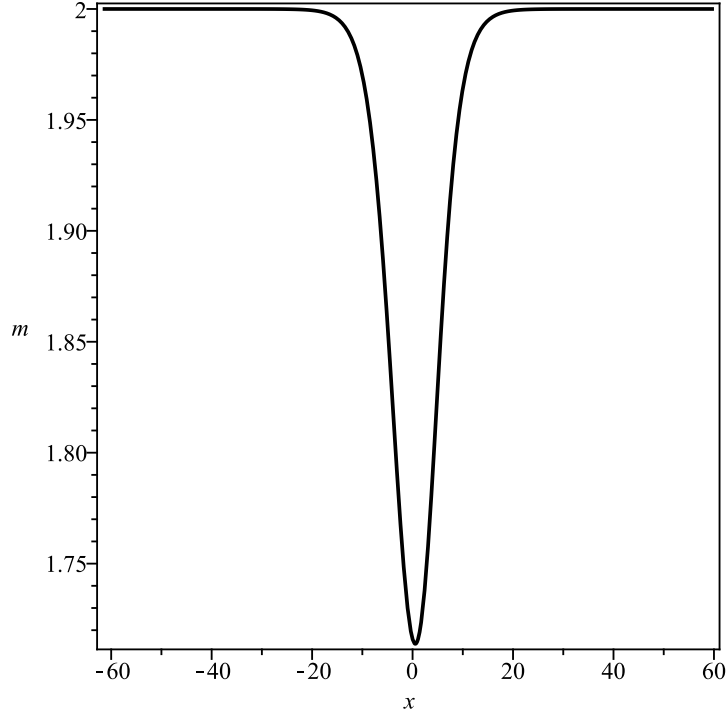}
\end{center}
\caption{One soliton profile, $m_0=2$, $ \kappa_1=0.2$.}
\label{one sol figure}
\end{figure}

The extremum (minimum) of $m$ occurs when \[\xi_1 = \frac{1}{4\kappa_1}\ln\left(\frac{1-m_0\kappa_1}{1+m_0\kappa_1}\right).\]
This is a constant value, e.g. the soliton moves with a velocity $ \frac{f(\kappa_1)}{2\kappa_1}$ that depends on the dispersion law (i.e. the chosen equation from the hierarchy). The profile of the dark soliton is given on Fig. \ref{one sol figure}.

\section{Example: Two soliton solution}

In the case of two discrete eigenvalues we compute

\begin{equation}\varphi_+(y,t, \frac{-i}{m_0}) = e^{\frac{y}{m_0}}\frac{1+\nu_1e^{-2\kappa_1 \xi_1}+\nu_2e^{-2\kappa_2 \xi_2}+ \left(\frac{\kappa_1-\kappa_2}{\kappa_1 +\kappa_2} \right)^2 \nu_1 \nu_2 e^{-2\kappa_1 \xi_1-2\kappa_2 \xi_2}}{1+e^{-2\kappa_1 \xi_1}+e^{-2\kappa_2 \xi_2}+ \left(\frac{\kappa_1-\kappa_2}{\kappa_1 +\kappa_2} \right)^2 e^{-2\kappa_1 \xi_1-2\kappa_2 \xi_2}}  \end{equation} where the following notation is utilized: $$ \nu_j = \frac{\frac{1}{m_0}-\kappa_j}{\frac{1}{m_0}+\kappa_j},\qquad j=1,2. $$
From (\ref{sol1}) and (\ref{sol2}) we obtain the two-soliton solutions:

\begin{eqnarray}
x(y,t)&=&\frac{2y}{m_0}+ 2\ln \frac{\Delta_1}{\Delta_2} \\
m(y,t) &=& \frac{m_0}{1 + \frac{m_0\Delta_3}{\Delta_1 \Delta_2}}.
\end{eqnarray}

where the following notations are used:

\begin{eqnarray}
\Delta_1(y,t)&=&1+e^{-2\kappa_1 \xi_1}+e^{-2\kappa_2 \xi_2}+ \left(\frac{\kappa_1-\kappa_2}{\kappa_1 +\kappa_2} \right)^2 e^{-2\kappa_1 \xi_1-2\kappa_2 \xi_2} \nonumber \\
\Delta_2(y,t) &=&1+\nu_1e^{-2\kappa_1 \xi_1}+\nu_2e^{-2\kappa_2 \xi_2}+ \left(\frac{\kappa_1-\kappa_2}{\kappa_1 +\kappa_2} \right)^2 \nu_1 \nu_2 e^{-2\kappa_1 \xi_1-2\kappa_2 \xi_2} . \nonumber\\  \Delta_3(y,t) &=& \frac{4\kappa_1 ^2}{m_0^{-1}+\kappa_1}e^{-2\kappa_1 \xi_1}+\frac{4\kappa_2 ^2}{m_0^{-1}+\kappa_2}e^{-2\kappa_2 \xi_2}\nonumber \\
&+&\frac{8(\kappa_1 -\kappa_2)^2}{m_0(m_0^{-1}+\kappa_1)(m_0^{-1}+\kappa_2)}e^{-2\kappa_1 \xi_1-2\kappa_2 \xi_2} \nonumber \\
&+&  \frac{4\kappa_2 ^2 \nu_1}{m_0^{-1}+\kappa_2}\left(\frac{\kappa_1-\kappa_2}{\kappa_1 +\kappa_2} \right)^2 e^{-4\kappa_1 \xi_1 - 2 \kappa_2 \xi_2} \nonumber \\
&+&  \frac{4\kappa_1 ^2 \nu_2}{m_0^{-1}+\kappa_1}\left(\frac{\kappa_1-\kappa_2}{\kappa_1 +\kappa_2} \right)^2 e^{-2\kappa_1 \xi_1 - 4 \kappa_2 \xi_2}.
\end{eqnarray}

\begin{figure}
\includegraphics[width=0.31\textwidth]{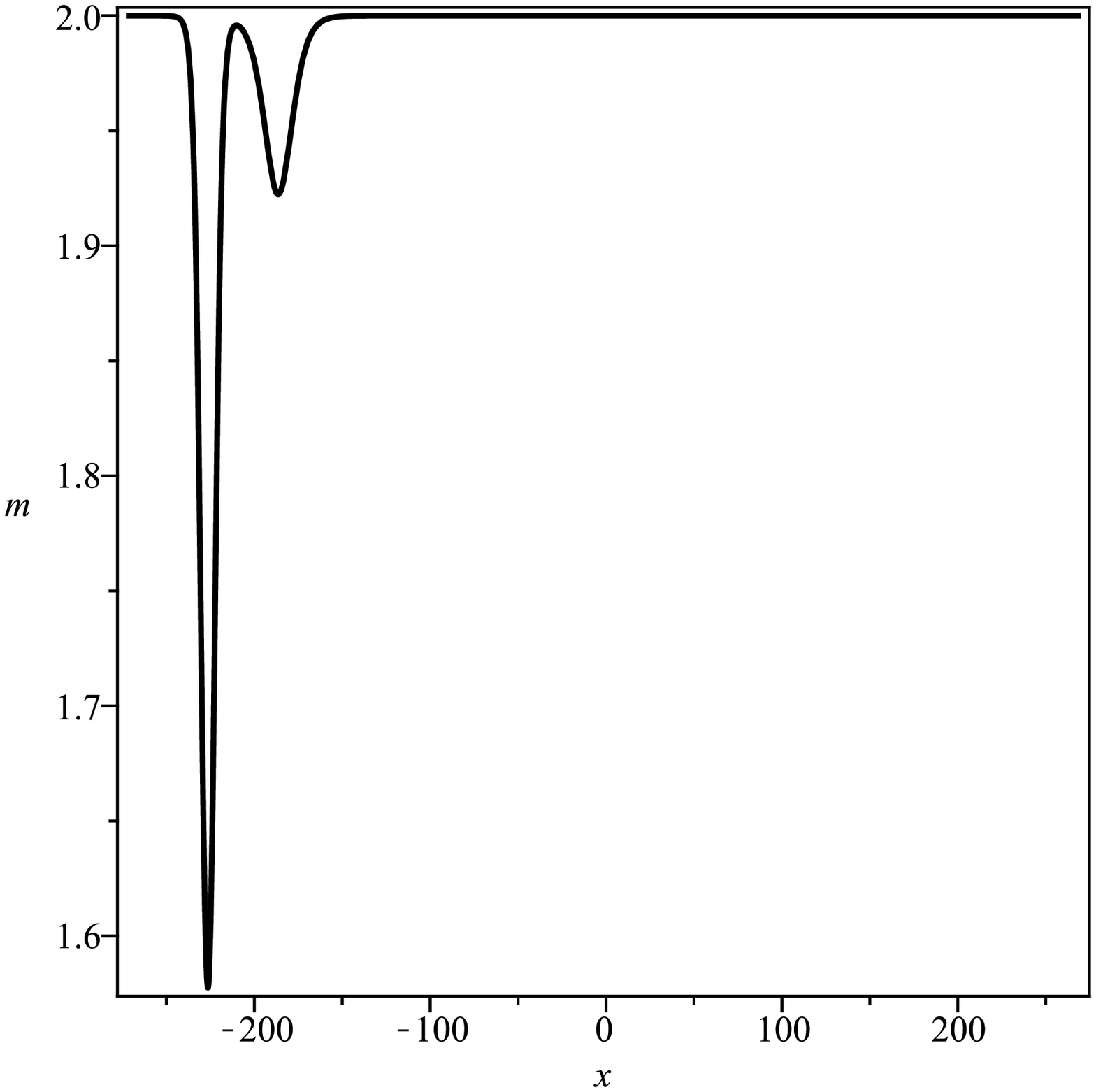}
\includegraphics[width=0.31\textwidth]{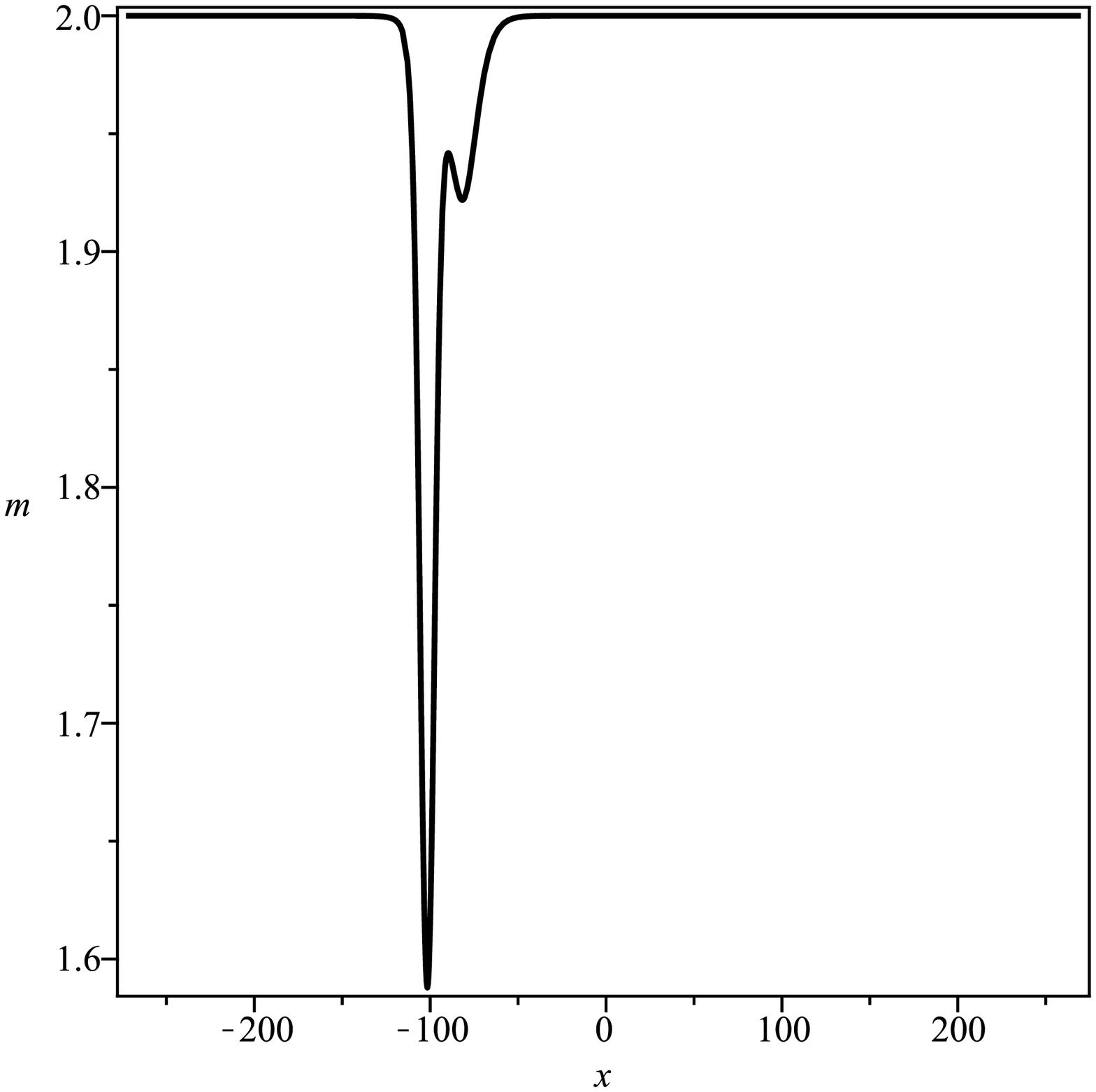}
\includegraphics[width=0.31\textwidth]{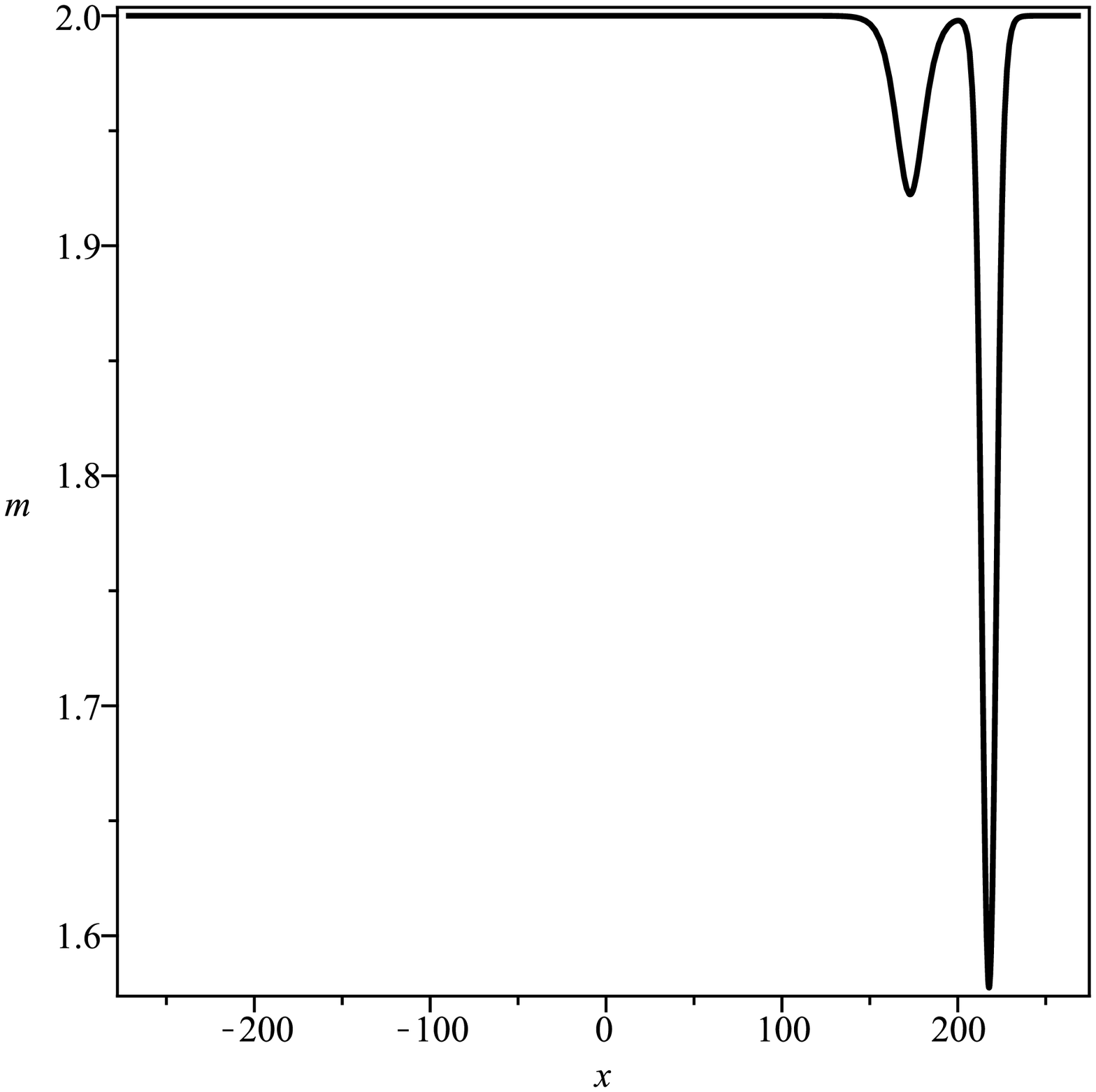}
\caption{Snapshots of the two (dark) soliton solution of the Qiao equation (\ref{Q}),
for three values of $t:$ $-30$, $-12$ and $30$. The other parameters are $m_0=2$, $\kappa_1=0.1$, $\kappa_2=0.25$.  } \label{two sol figure}
\end{figure}

The interaction of two dark solitons is illustrated on Fig. \ref{two sol figure}.

\section{Conclusions}

In this paper we demonstrated how the spectral problem for the
Qiao's hierarchy can be reduced to the one for the standard
Schr\"odinger operator and hence the soliton solutions ('dark'
solitons) can be obtained in a straightforward manner. This
necessitates constant boundary conditions for the solution and
also a restriction on the discrete eigenvalues $0<\kappa_n <
m_0^{-1}$. It is interesting what happens to the solutions if this
condition is violated. Based on the similarity with Camassa-Holm
equation it is likely that there are breaking waves present in
this case. Moreover, the equation (\ref{Q}) has a conservation law
in the form $$ X_x(x,t) m(X,t)=m(x,0)$$ where  $X$  is the
solution of
$$X_t(x,t)=u^2(X,t)-u_x^2(X,t), \qquad X(x,0)=x.$$ It is likely
that this conservation law will play an essential role in the
study of the wellposedness, existence and breaking of the
solutions.

\section{Acknowledgments}

The authors are indebted to Prof. A. Constantin for valuable
discussions and to an anonymous referee for some important
suggestions. This material is based upon works supported by the
Science Foundation Ireland (SFI), under Grant No. 09/RFP/MTH2144.

\end{document}